# Piezo-driven sample rotation system with ultra-low electron temperature

Pengjie Wang,[1] Ke Huang,[1] Jian Sun,[1] Jingjin Hu,[1] Hailong Fu,[2] Xi Lin[1,3,4*]

[1] *International Center for Quantum Materials, Peking University, Beijing 100871, China*
[2] *Department of Physics, The Pennsylvania State University, University Park, Pennsylvania 16802, USA*
[3] *CAS Center for Excellence in Topological Quantum Computation, University of Chinese Academy of Sciences, Beijing 100190, China*
[4] *Collaborative Innovation Center of Quantum Matter, Beijing 100871, China*

*xilin@pku.edu.cn

**Abstract**

Piezo-driven rotator is convenient for tilted magnetic field experiments due to its precise angle control. However, the rotator itself and the sample mounted on it are difficult to be cooled down because of extra heat leaks and presumably bad thermal contacts from the piezo. Here we report a piezo-driven sample rotation system designed for ultra-low temperature environment. The sample, as well as the rotating sample holder, can be cooled to as low as 25 mK by customized thermal links and thermal contacts. More importantly, the electron temperature in the electrical transport measurements can also be cooled down to 25 mK with the help of home-made filters. To demonstrate the application of our rotation system at ultra-low electron temperature, a measurement revealing tilt-induced localization and delocalization in the second Landau level of two-dimensional electron gas is provided.

## I. Introduction

The demand for a technique of changing the magnetic field in both magnitude and direction at low temperatures has been boosting in the past decades. The alteration of the magnetic field in the electrical transport measurements has played an indispensable part in the breakthrough of fractional quantum Hall (FQH) effect,[1] giant magnetoresistance[2,3] and quantum anomalous Hall effect,[4] etc. The in-situ alteration of the magnetic field direction in a low-temperature environment provided further opportunity for the research of exotic but fragile quantum states. For examples, in the search of Majorana fermion in semiconductor wires, tilted field experiments in a dilution refrigerator have been used to check the dependence of the zero-bias peak on the external field orientation;[5–7] in the two-dimensional electron gas (2DEG), rotating sample holder at ultra-low temperature opens up opportunities in pursuit of spin configurations,[8,9] subband degree of freedom,[10] and non-Abelian FQH candidates;[11–16] in the quantum dot system, the in-plane magnetic field can tune the Zeeman energy of the spin levels, which is a straightforward illustration for quantum two-level systems.[17,18]

The in-situ alteration of the magnetic field direction can be realized in a few methods, such as sample rotation system,[19] vector magnets[20] and magnet rotation system.[21] However, the latter two methods require special designs, such as the application of the Helmholtz coil, demanding a higher cost and more space compared to solenoid magnet. What's more, in the commercially available vector magnets, although the largest magnetic field can typically reach 9 T, it can achieve 9 T only in one direction and less than 3 T in other two directions.

The sample rotation system proves to be an easier method to alter the in-plane and out-of-plane components in a high magnetic field, which has been applied to achieve a total magnetic field of 45 T with a continuous rotation of 360°.[22,23] To rotate the sample in vacuum, the widely applied method is based on the worm-gear mechanism,[24,25] but it has difficulties accommodating with the commercially available millikelvin environment due to the remarkable frictional heat dissipation. With specified bearings made from sapphire vee jewels and matching pivots, Palm and Murphy[26] have reduced the frictional heating and have shown a rotation system with a base temperature down to 30 mK. Another approach exploiting the pressurized liquid $^3$He has been realized down to 8 mK, but it requires extra gas handling system and more complicated sample cell.[27] In the 2000s, the successful application of the piezo-driven rotator[28] brought out a solution for higher resolution and much more precise step control in ultra-low temperature environment. In 2010, Yeoh *et al.*[29] reported an integration of a commercial rotator from attocube with a wet dilution refrigerator below 100 mK, achieving a continuous *in-situ* rotation of more than 100°.

Although the piezo-based rotation system with ultra-low temperature has been technically feasible, lowering the temperature of electrons is more important. The commercial quantum annealing machine from D-Wave has shown its strength in quantum optimization[30] and quantum simulation,[31,32] where qubits manifest their quantum properties of entanglement and tunneling only at ultra-low temperatures.[33] The quantum computation based on superconducting qubits also requires a low electron temperature to stabilize the quantum states.[34,35] For electrical transport measurements which need to know the temperature precisely, lowering the electron temperature is even more crucial.[36] Meanwhile, an ultra-low electron temperature compatible piezo-driven rotation system is technically valuable in realizing scanning tunneling microscopy (STM)[37] and atomic force microscopy (AFM)[38] in dilution refrigerators because the piezo-driven system in STM and AFM is exactly one of the hurdles in their cooling in dilution refrigerators. At ultra-low temperatures, the electrons in a sample are typically at a higher temperature than the environment because of the weak electron-phonon couplings. Even a sample locates in a 10 mK environment and is well thermalized, the temperature of the conducting electrons in it can be higher than 100 mK due to extra heating along the leads and from the environment. Cooling the electrons is challenging below 100 mK, and a piezo-driven rotation system makes it more strenuous.

In this work, we report our design of piezo-driven sample rotation system based on a commercial piezo-driven rotator (Model ANR101/LT/RES from attocube systems AG) and a sample-exchange cryogenic probe in a dilution refrigerator (Model CF-CS81-600 from Leiden Cryogenics BV). This piezo-driven rotation system can reach a low electron temperature down to 25 mK. The *in-situ* sample holder can rotate through a total angle range of 98° from less than 0° to more than 90° with an angle readout resolution of ±0.02° and stability within ±0.05°. We present our design details on the mechanical assembly, leads for transport measurements, and filtering for lowering the electron temperature. We also demonstrate this system's performance with a transport measurement in a GaAs/AlGaAs 2DEG.

## II. Mechanical Assembly

The piezo-driven rotator is mechanically mounted to the rotator holder (item #1 in Fig. 1 (a)) made of oxygen-free high-conductivity (OFHC) copper, which connects to an OFHC copper cold-finger from the mixing chamber (MC) plate of the cryogenic probe, as shown in Fig. 1(a). It should be noted that good electrolytic copper with oxygen is a better thermal conductor at low temperatures, but here we demonstrate that OFHC copper also works for our application. Slots are cut on the rotator holder to suppress eddy current in magnetic fields. All the screws, nuts and gaskets used in our setup are made of 304 stainless steel. A schematic view of the thermal and wiring connections for the sample is shown in Fig. 1(b).

### a) Thermalization of the rotator

The rotator base (#2) is directly cooled by the mechanical connection to the rotator holder (#1), as shown in the back view of Fig. 1(a). However, the thermal connection between the rotator base (#2) and the rotation stage (#3) is too weak at dilution temperatures due to the ceramic tube for the slip-stick procedure. When the MC plate is cooled down to 25 mK, the temperature of the sample carrier can still be as high as 42 mK in a well-designed experiment, if it is only cooled by mechanical support and wirings without extra thermal links.[39] For a low electron temperature environment, the effective cooling of the rotation stage (#3) is necessary. Otherwise, the relatively hot BeCu rotation stage (#3) will be a continuous heat source for the sample even though there is plastic insulation, such as the sample base (#4) in Fig. 1(a), between them.

As shown in the back and bottom view of Fig. 1(a), a pair of copper-braid with a diameter of 1 mm is used as thermal link (#5) to cool down the rotation stage (#3) at low temperature. One end of the copper braid is argon-arc welded to an OFHC copper connector, which is fixed to the rotation stage (#3) by two M2.5 screws. The copper braid passes through the 2 mm diameter center hole of the rotator, minimizing the torque needed in a rotation process. The other end of the copper braid is glued to another OFHC copper connector by silver paint (item no. 05002-AB from SPI supplies), which is fixed to the rotator holder (#1), as shown in the back view of Fig. 1(a). Welding is avoided at this end of the copper braid because the copper braid is as short as 60 mm thus the heating from the welding process could damage the rotator.

### b) Sample base

The sample base (#4) is made of polytetrafluoroethylene (PTFE) and shaped to fit the rotation stage (#3) and sample carrier (#6), as shown in the front and back views of Fig. 1(a). 24 female pins are glued in the corresponding 24 holes by Stycast 2850FT epoxy to fit the sample carrier (#6). PTFE serves as a good thermal insulator in millikelvins and can also minimize the eddy current heating in magnetic fields, which is an important trick to achieve ultra-low electron temperature. The female and male pins for our sample rotator are gold-plated brass without nickel flashing. The total heat capacity of the PTFE is much lower than that of copper at the relevant temperature range, so the application of PTFE here will also shorten the thermal equilibrium time at low temperature.

### c) Sample holder

The sample holder consists of two main parts: sample carrier (#6) and L-shape connector (#7). These two parts are connected to each other by two pairs of copper-braid with a diameter of 1.5 mm as thermal link (#8). The connection joints are argon-arc welded. The L-shape connector (#7) is made of OFHC copper and fixed to the rotator holder (#1) as shown in the front view of Fig. 1(a). The sample carrier (#6) is a 23 mm × 17 mm square made of OFHC copper with a thickness of 1 mm. The sample holder is gold-plated after welding, so the thermal links (#8) are partly covered with a thin layer of gold, but the torques induced by the gold plating do not prevent the rotating of the rotator. Two slits are cut to decrease the eddy current in magnetic fields. There are 24 through holes with a diameter of 1.7 mm and two through holes with a diameter of 3.2 mm on the sample carrier (#6). Gold-plated electrical pins are fixed to the sample carrier in the 24 smaller through holes respectively by Stycast 2850FT epoxy. The larger spare holes can be used for other types of sample base without a rotator. The space for the samples is 14 mm × 14 mm, suitable for a maximum of four samples with a size of 5 mm × 5 mm in one cool-down. Samples can be glued onto the sample carrier (#6) by a thin layer of Ge-varnish, as shown in Fig. 1(b). Gold wires ($d$ = 0.001 inch, from Semiconductor Packaging Materials Inc.) can be bonded to connect the samples and the electrical pins.

### d) Red light-emitting diode (LED)

A red LED (#9) is mounted on the rotator holder (#1) to illuminate the samples at cryogenic temperature, as shown in Fig. 1(a). In the research of FQH effect, the 2DEG in most GaAs samples have higher mobility and higher quality after appropriate exposure in red LED at appropriate temperature.[40,41]

### e) Angle determination

The angle of the rotation could be determined by a positioner, i.e. a 3-terminal film resistor, embedded in the rotator, as shown in Fig. 2(a). Lead 1 and lead 3 are fixed while lead 2 rotates with the rotation stage (#3). In Fig. 2(b), the side view of the rotator is shown and the touching tip of the positioner (lead 2) is labeled. The resistance of the whole resistor (between lead 1 and lead 3) is around 20 k$\Omega$ for 315°. An ac voltage of 2 mV at 133 Hz is applied between lead 1 and lead 3, and lead 2 is connected to a preamplifier and a lock-in amplifier, as a probe for the angle. If the film resistor is homogeneous, the rotation angle is proportional to the readout voltage. This has been demonstrated in the pioneer works.[29,42] In our system, the readout resolution of the angle is within ±0.02°, similar to the previous results.[29] The rotation angle is defined as $\theta$ in Fig. 2(c). $\theta$ = 0° means the total magnetic field is perpendicular to the sample plane, and $\theta$ = 90° means the total magnetic field is parallel.

## III. Leads for Transport Measurements

One advantage of the piezo-driven rotation system is the electrical lead-only design, so that people do not need to introduce any mechanical links or gas lines from room temperature down to millikelvin temperature. An illustration of our wirings for transport measurements is provided in this section.

### a) Leads for the rotator

Five wires (#10) are used between room temperature connectors and the rotator at the MC plate, two of which provide the drive voltage to the piezo, and the rest three are used for the angular position resistor. The two drive lines require a very low resistance (< 5 Ω) to provide a sharp voltage drop on the piezo, allowing for slip-stick rotation. Nevertheless, low-resistance copper wires between room temperature and the MC plate are not applicable because too much heat would be brought down and warm up the millikelvin environment of the MC plate. In our first design, a break was added at the 3K plate and copper wires ($d$ = 0.1 mm, shielded by braided CuNi, supplied by Leiden Cryogenics BV) were only used from room temperature connector down to the 3K plate while NbTi superconducting wires ($d$ = 0.1 mm, cladded with Cu/Ni alloy 70/30, shielded by braided CuNi, supplied by Leiden Cryogenics BV) were used between the 3K plate and the MC plate. When the temperature is below 10 K, NbTi turns to superconducting under zero field, which reduces the heat leak from the 3K plate to the MC plate. All the wires were thermalized at each plate of the probe that they went through.

This setup worked well at a low probe temperature down to 14 mK with wire resistance less than 1.0 Ω per wire. However, at room temperature, the total resistance between the room temperature connector and the rotator was as high as 25 Ω per wire, which was too large to trigger the piezo for tests. Therefore, in our recent application, we replaced the combination set of copper wire and traditional superconducting wire with 2% nickel alloyed copper wire ($d$ = 0.2 mm, shielded by braided CuNi, supplied by Leiden Cryogenics BV, also named nickel alloy 30 wire).

Both designs can achieve the 25 mK electron temperature environment. The advantage of 2% nickel alloyed copper wires is that they are alloys so that the resistivity ratio stays constant with temperature, and the thermal conduction is low enough for this application. The resistance per unit length of the wire is 1.7 Ω/m. With this, the wires could go down directly to the MC plate without adding a break and one can easily check the rotation system at room temperature.

For the positioner, i.e. the 3-terminal resistor in the rotator, we used the same type of wires as the drive lines to lower down the wire resistance.

b) **Leads for the sample base**

The leads for the sample base can be separated into three parts: wirings from room temperature to the MC plate, wirings from the MC plate to the filters and wirings from the filters to the sample base on the rotator. For the wirings from room temperature to the MC plate, a braided CuNi shielded cable with 24 CuNi inner wires (36 AWG, spec no. C71500 ALLOY from Calmont Wire & Cable) is used. There is a graphite coating between the dielectric and outer shield to reduce triboelectric noise. The cable is thermalized at each plate of the probe. For the stage between the MC plate and silver-epoxy filters (filters are introduced in the next section), 24 coax cables with a 36 AWG CuNi inner wire and braided CuNi outer shield are used. The silver-epoxy filters are located on a plate below the MC plate. For the third part, i.e. from the silver-epoxy filters to the sample base, there are 12 pairs of twisted enameled copper wires ($d$ = 0.15 mm, spec no. CUL 100/0,15 from Block Transformatoren GmbH) with a set of switch connectors mounted on a cold plate underneath the silver-epoxy filters. The last part of twisted enameled copper wires (#11) is thermalized with silver paint (item no. 05002-AB from SPI supplies) to the cold plate, as illustrated in Fig. 1(b). The thermalized wires would also help cool down the sample effectively.

c) **Leads for the red light-emitting diode (LED)**

Two coax cables, i.e. CuNi wires (36 AWG, spec no. C71500 ALLOY) with CuNi outer shield from Calmont Wire & Cable, are used from two room temperature LEMO connectors down to two MMCX connectors at the MC plate. Both cables are thermalized at each stage of the probe. The same type of enameled copper wires as those used for sample base are used from MC plate to the LED, labeled as item #12 in Fig. 1(a). The current for the LED is typically no larger than 10 mA at cryogenic temperature.

**IV. Filtering for Low Electron Temperature**

In ultra-low temperature transport measurements, weak electron-phonon interaction and high-frequency noise from the measurement instruments and environment usually cause the electron temperature to elevate from the lattice temperature. Therefore, effective filtering for the sample is indispensable for electrical transports that need precise information of the temperature. In this section, we introduce our microwave silver-epoxy filters at the low-temperature stage and kHz RC filters at room temperature. An Agilent E5071C ENA vector network analyzer has been used to record the performance of the filters.

### a) Silver-epoxy filter

There are a few different approaches to construct cryogenic microwave filters. The most traditional method is to take the advantage of the skin effect, which has been applied by metallic powder filters,[43–46] Thermocoax,[47] silver-epoxy filters,[48] etc. Among them, the silver-epoxy filter integrates several pros together, i.e. smaller size, better performance, and convenient thermalization. In our silver-epoxy filters, 3.7 m of enameled copper wire ($d = 0.1$ mm, spec no. CUL 100/0,10 from Block Transformatoren GmbH) is wound around a 1.0 mm thick Ag wire to form a 7-layer coil of 10 mm length. We keep coating the silver epoxy (Epo-tek E4110) onto the coil to make sure it is fully covered. Standard MMCX connectors are soldered at both ends and filled with Stycast 2850FT epoxy for electrical isolation to the shielding. Finally, we glue all parts together with silver epoxy to ensure good thermalization and electrical connection. Our silver-epoxy filters are shown in Fig. 3(a). A detailed study of the silver-epoxy filter can be found in the work by Scheller *et al.*[48]

### b) RC filter

We have designed a 3$^{rd}$-order RC filter to lower the noise at kHz range, while MHz and GHz radiation cannot be attenuated effectively due to stray capacitances and resistances of the components. The circuit diagram of the RC filters is shown in Fig. 3(b), designed for a cut-off frequency of 22 kHz. The resistors are 510 Ω, 820 Ω and 1500 Ω while the capacitors are 4.7 nF, 2.2 nF and 1.1 nF for each stage respectively. We solder all components onto a printed circuit board, which is mounted in an aluminum shield with BNC connectors at both ends, as shown in Fig. 3(c). RC filters are used to connect all the wires individually behind the breakout box at room temperature. The future improvement could be done by adapting the RC filters at base temperature to suppress the thermal noise from the resistors.[49]

### c) Performance of filters

In Fig. 3(d), we show the attenuation of the silver-epoxy filter (black line), RC filter (red line) and their combination in series (blue line) versus the frequency at room temperature. The silver-epoxy filter has an attenuation of more than 90 dB, reaching the noise floor when the frequency is larger than 600 MHz. The noise floor is the signal created from the sum of all the noise sources and undesirable signals. The performance of our silver-epoxy filter is better than a 3-meter Thermocoax cable in reference [49]. For the RC filter, the weakening of the attenuation in MHz and GHz range is due to the existence of the stray capacitances and resistances. The combination of the silver-epoxy filter and RC filter in series provides a good attenuation reaching more than 90 dB at 100 MHz. Similar filtering with an effective cooldown of the sample could reach an electron temperature down to 15 mK without a rotator.[49]

## V. Estimate of Electron Temperature

The estimate of electron temperature is based on a high-quality GaAs/AlGaAs sample with van der Pauw geometry (sitting on top of the sample carrier in Fig. 1(a)) mounting on the sample rotation system. The density is $3.2\times10^{11}$ cm$^{-2}$ and the mobility is $2.8\times10^{7}$ cm$^2$ V$^{-1}$ s$^{-1}$. The quantum well width of the sample in the z-axis is 28 nm, so electrons can only move freely in xy-plane at low temperatures. The sample was illuminated with a red LED at 4.5 K with 15 μA for 1 hour before the measurements were taken. Standard lock-in technique was applied with an ac excitation of no more than 8 nA and a frequency of 17 Hz.

In Fig. 4, the longitudinal resistance $R_{xx}$ and Hall resistance $R_{xy}$ as a function of total magnetic field $B_{\text{total}}$ in the second Landau level at 25 mK are shown. The exactly quantized 7/3, 5/2 and 8/3 FQH states, whose energy gaps are in the order of 100 mK or less, only appear in high-mobility 2DEG at ultra-low temperature. The emergences of all eight fragile reentrant integer quantum Hall (RIQH) states in the second Landau level are more difficult than the 7/3, 5/2 and 8/3 FQH states. Moreover, three RIQH states labeled as R2a, R2c, and R3a in Fig. 4 are exactly quantized, which is usually expected at electron temperature around 20 mK.[49–54] All the above indicate the effective cooldown of the electron temperature in our sample rotation system.

To determine the electron temperature quantitatively, the activated behavior of quantum Hall states is exploited. In the thermally activated regime, the activation energy gap of a quantum Hall state could be identified by the temperature dependence of longitudinal resistance minimum, i.e. $R_{xx} \propto \exp(-\Delta/2k_\text{B}T)$. Because the quantum Hall effect origins only from electrons, the temperature affecting transport behaviors is indeed the electron temperature. Therefore, the electron temperature could be inspected by an Arrhenius plot ($\ln R_{xx} - T^{-1}$). If the electron temperature of a sample is higher than the environmental temperature at the low-temperature limit, the logarithm of $R_{xx}$ will deviate from the linear relation of the activated energy gap when using the temperatures from the thermometer in the Arrhenius plot.[49]

A thermometer mounted in a dilution refrigerator measured the environmental temperature and it was electrically isolated from the 2DEG in samples. The thermometer we used in the measurement was a ruthenium oxide resistor calibrated by a noise thermometer. We installed the resistive thermometer at the magnetic field compensating region below the MC plate of the probe to minimize the magnetoresistance. In the same cooldown, a cerium magnesium nitrate thermometer was used to double check the temperature of the resistive thermometer. The thermometry for the low temperature environment is reliable down to 7.5 mK in our dilution refrigerator.

Based on the energy gap of a fragile 7/2 FQH state[10,14,50,55–57] and environmental temperature thermometry, here we demonstrate that the electron temperature in this rotation system is effectively cooled down to 25 mK. In Fig. 5(a), the traces of the upper spin branch of the second Landau level taken at different temperatures are shown. The dips at the filling factor of 7/2 show that the 7/2 FQH state is appearing. The lower the dip, the lower the electron temperature. We fixed the magnetic field at the center of the 7/2 FQH state (3.798 T) to get rid of magnetization warming or demagnetization cooling during a magnetic field sweep. Then we changed the probe temperature and waited for the probe temperature (red line) and the longitudinal resistance $R_{xx}$ of the 7/2 FQH state (black line) to reach equilibrium, as shown in Fig. 5(b). The temperatures (red line) shown in Fig. 5(b) are from the resistive thermometer installed at the field-compensating region. Averaging over the stable longitudinal resistance at specific temperatures, the averaged longitudinal resistance $R_{xx}$ as a function of the temperature $T$ is plotted in Fig. 5(c). From a linear fitting of the data in the low-temperature limit, the energy gap of the 7/2 FQH state is 119.8 mK. The resistance of the 7/2 FQH state at 25 mK still obeys the linear relation of the energy gap in Fig. 5(c), suggesting the electron temperature of the sample should still be the same as that the thermometer indicates. The measurement has been repeated multiple times independently to check its reproducibility.

From Fig. 5(b), we can also conclude that the thermal equilibrium time of our system is within 20 mins for a 5 mK temperature step above 30 mK. The longer thermal equilibrium time from 30 mK to 25 mK results from the longer thermal equilibrium time of the probe temperature.

### VI. Application of Sample Rotation System

Here, we provide a study on FQH effect in the same sample of Fig. 4, as an example of the application of our sample rotation system. In this study, the localization of the FQH liquid and delocalization of the solid phases in the second Landau level is probed with a tilted field at 25 mK electron temperature.

For 2DEG exposed in a given magnetic field $B_{total}$, the best way to calibrate the exact titled angle $\theta$ is its cosine relation with the Hall resistance at different tilted angles, i.e. $R_{xy} \propto B_{total} \cos\theta$, so the Hall resistance $R_{xy}$ is 0 Ω when the tilted angle $\theta$ is 90°. In our test, the relative readout error $\pm\Delta R/R$ for Hall resistance is ±0.03%, which can correspond to an angle accuracy of 0.02° at 90° in our system.

The stability of the tilted angle is confirmed by measuring the same Hall trace multiple times after thermal cycle and magnetic field sweeping. If there is slow angle drift, then the slope of Hall resistance or plateau position in the magnetic field is supposed to vary with time or vary after measurement sequence.

We set the rotation angle at 57.4°, an arbitrary angle, and did repeated measurements of the Hall traces after field sweeps, thermal cycles, or with a total time interval of 17 hours. As expected, we could not find any difference between the traces. Consider that the stability and readout error of the magnetic field is less than ±0.001 T in our system, the stability of the tilted angle should be within ±0.05°.

The Hall resistance as a function of total magnetic field $B_{total}$ at different tilted angles are taken at 25 mK and plotted in Fig. 6(a). The larger the tilted angle, the lower the slope. Once we sweep the magnetic field to higher than 3.5 T, the RIQH states in the second Landau level develop. These states in the second Landau level are fragile with onset temperature lower than 55 mK.[52] Before this demonstration, only one tilted field measurement on the RIQH states in the second Landau level has been conducted in a hydrostatic liquid $^3$He driven sample rotator so far.[58]

In Fig. 6(b), the Hall traces in the lower spin branch of the second Landau level at different tilted angles and 25 mK are shown. A constant offset of 0.05 $h/e^2$ between every two traces is applied for clarity. With the tilted angle increasing, 2+4/5 FQH state and all four RIQH states are destroyed quickly while the 2+1/3 and 2+2/3 FQH states stay robust. After destroyed, the 2+4/5 FQH state does not re-emerge with further tilting the sample but evolves into the nearest integer quantum Hall state, while the RIQH states disappear. The localization behavior at filling factor 2+4/5 could be the evolution from FQH liquid ground state toward a collective pinned bubble state with tilted angle, while the delocalization of the RIQH states could be understood as a tilt-induced melting of bubble states.[58]

**VII. Summary**

In this work, we report our design of a low-electron-temperature sample rotation system based on an attocube rotator and a Leiden cryogen-free dilution refrigerator. Samples mounted on this system can reach a base electron temperature as low as 25 mK under a magnetic field up to 10 T with reasonably short thermal equilibrium time. Samples can be rotated *in-situ* through a total angle range of 98° with a readout resolution of ±0.02° and stability of ±0.05°. We show an application of our design in a study of tilt-induced localization and delocalization of 2DEG in the second Landau level, which has only been studied in a pressurized liquid $^3$He cell before.[58]

**Acknowledgements**

We thank Prof. Giorgio Frossati from Leiden Cryogenics B.V. and Dr. Markus Janotta from Attocube Systems AG for their supports on the dilution refrigerator and sample rotator. We thank Prof. Loren Pfeiffer and Dr. Ken West for the high-quality sample. We thank Yang Liu for fruitful discussion. This work is supported by the National Key Research and Development Program of China (Grant No. 2017YFA0303301 and Grant No. 2015CB921101), NSFC (Grant No. 11674009), and supported by the Strategic Priority Research Program of Chinese Academy of Sciences, Grant No. XDB28000000.

**Figures**

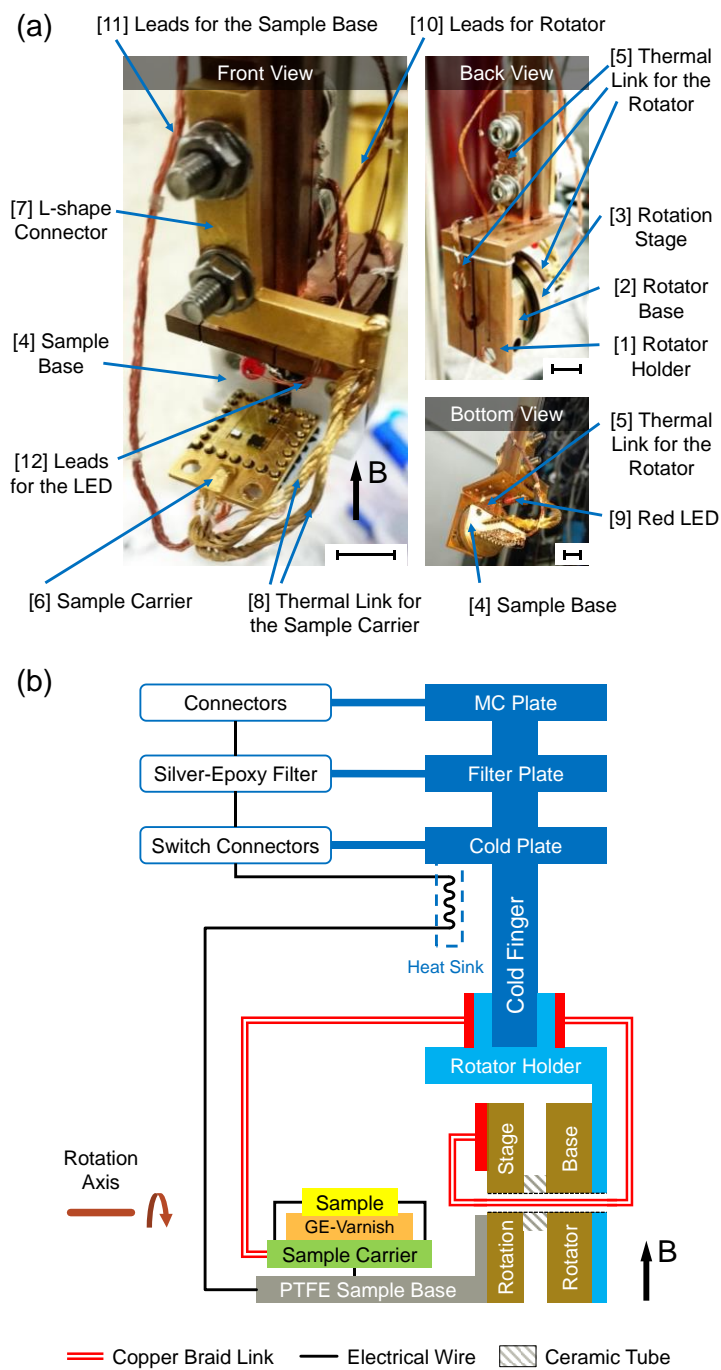

**Fig. 1** (a) Picture of rotator assembly: front, back and bottom views illustrating the assembly of rotator, wirings, and thermalization. The scale bars in all three views represent 10 mm. Because the pictures are taken with angles, the scale bars can be used as a reference only for the center of the pictures. (b) Schematic view of the thermal and wiring connections with a sample mounted. The sizes of parts are different from the real scale for clarity. The rotation axis and the external magnetic field direction are labeled.

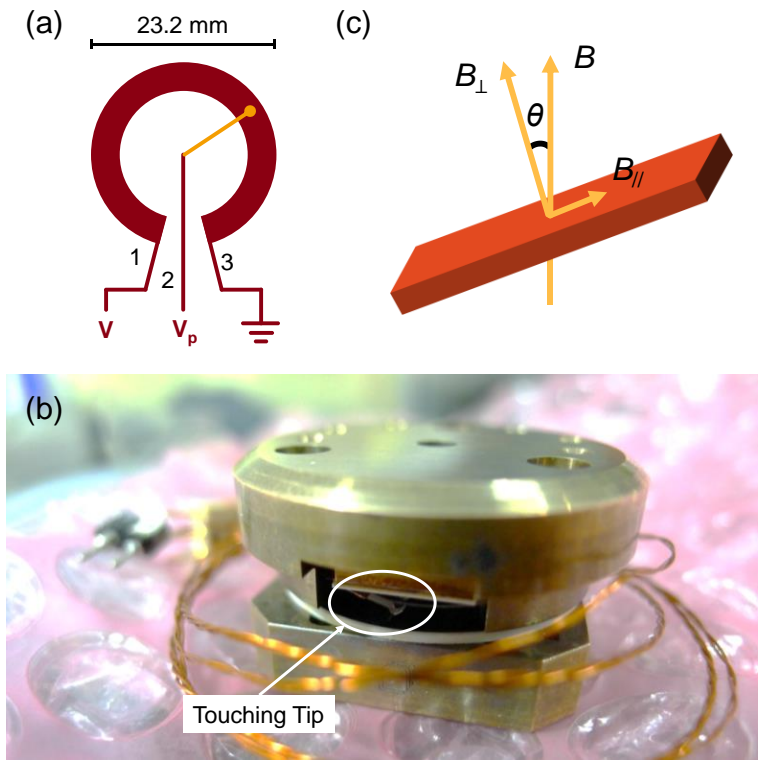

**Fig. 2** (a) Schematic view of the positioner located inside the sample rotator. (b) Side view of the sample rotator showing the touching tip of the positioner (Lead 2 in Fig. 2(a)). (c) Schematic view of a sample in a tilted field. Our system can rotate through a total angle range of 98° from less than 0° to more than 90°. The slightly negative angle helps to calibrate the perpendicular situation.

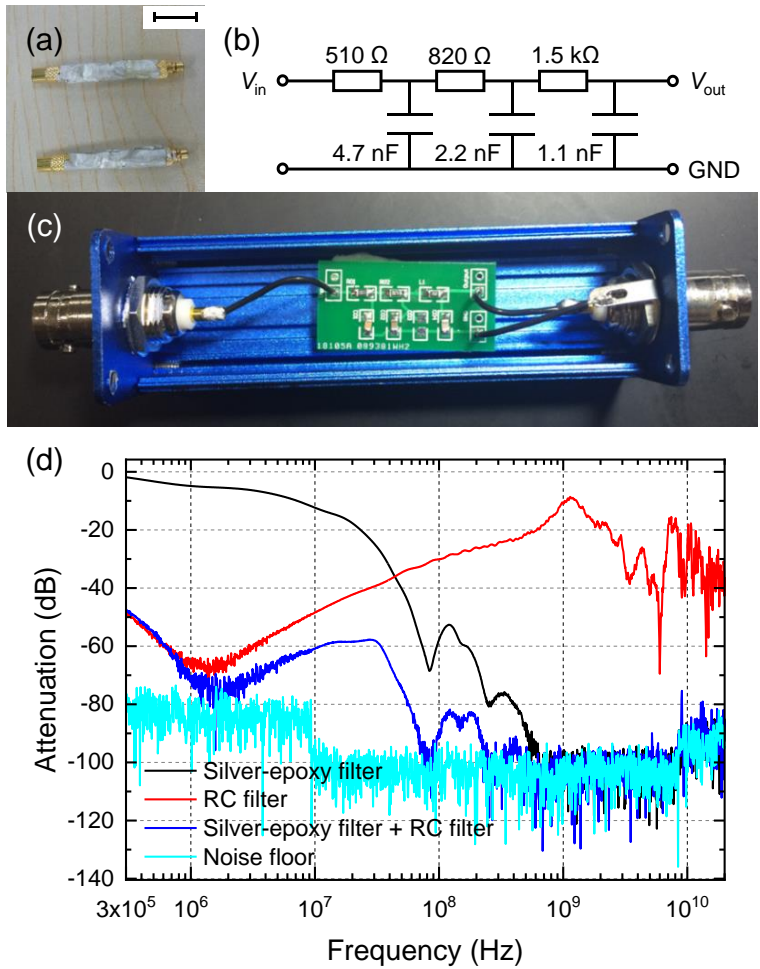

**Fig. 3** Filtering for lowering the electron temperature. (a) Home-made cryogenic silver-epoxy filters. The scale bar is 10 mm. (b) Layout of the room temperature 3$^{rd}$-order RC filter. (c) Home-made 3$^{rd}$-order RC filter. (d) Performance of individual home-made filters and their combination.

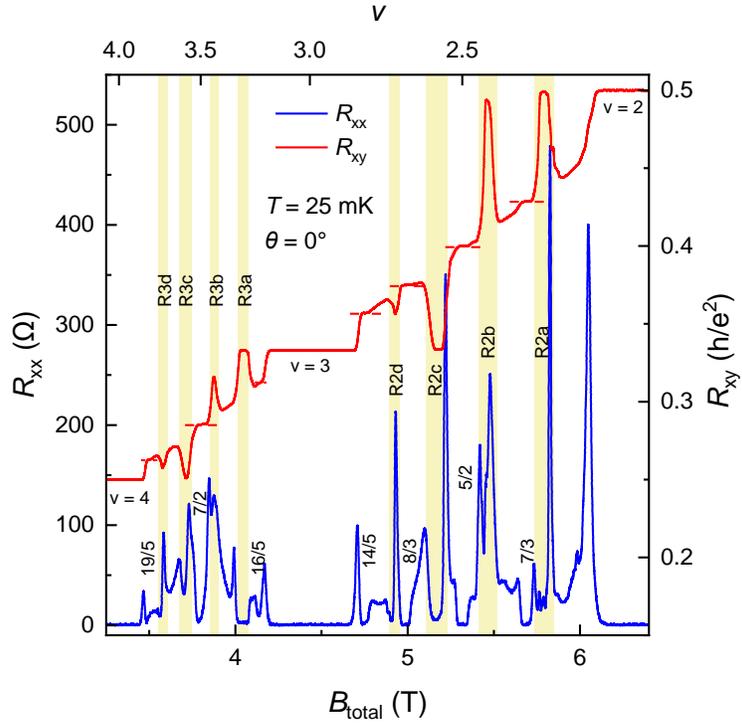

**Fig. 4** Longitudinal resistance (blue line) and Hall resistance (red line) traces in the second Landau level ($2 < v < 4$) taken at a probe temperature of 25 mK without tilting. Filling factors of the prominent FQH states are labeled. The yellow shaded regions represent the RIQH states, whose appearance indicates high sample quality and low electron temperature.

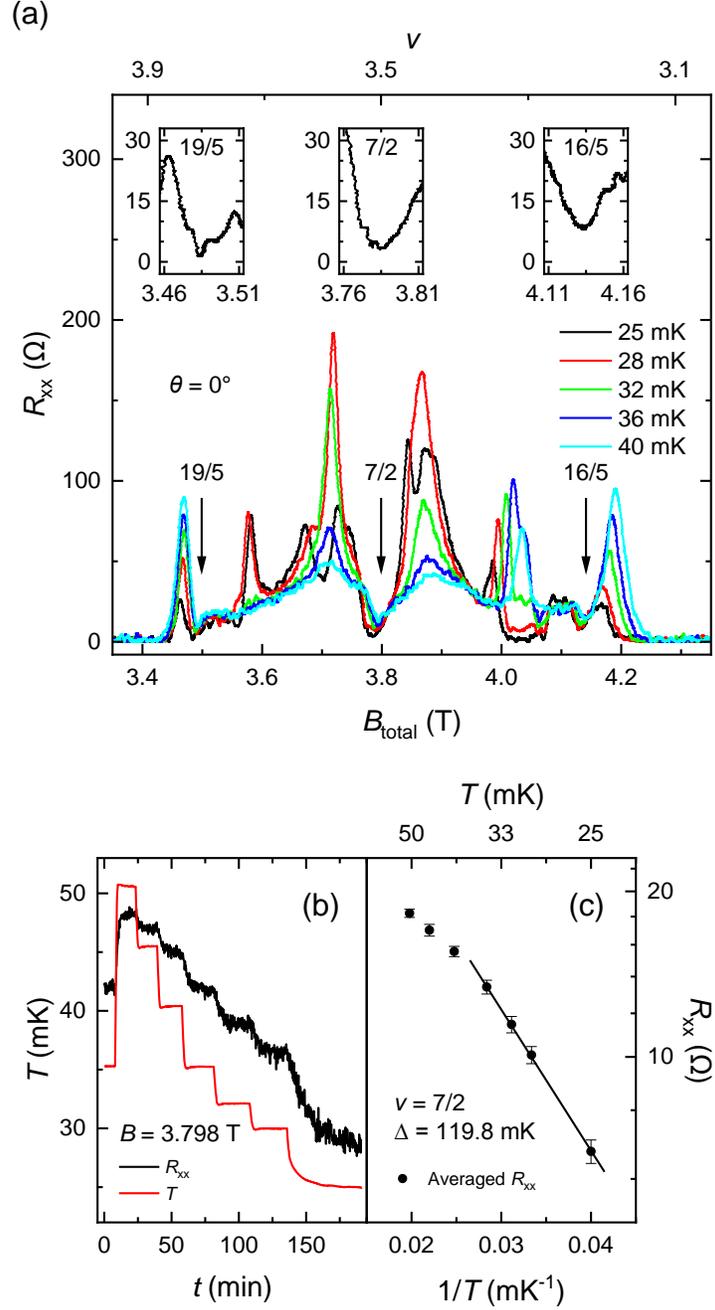

**Fig. 5** (a) Probe temperature dependence of the longitudinal resistance trace in the upper spin branch of the second Landau level ($3 < \nu < 4$) without tilting. Three prominent FQH states are marked with arrows. The zoomed in insets show the longitudinal resistance dips of 16/5, 7/2 and 19/5 FQH states at 25 mK. (b) Time elapse for the longitudinal resistance when the magnetic field $B$ is set at 3.798 T, where the center of the 7/2 FQH state locates. The short thermal equilibrium time shows good thermal equilibrium condition for the sample. (c) Arrhenius plot of the 7/2 FQH state. The black circle symbols are the averaged longitudinal resistance when the sample is at thermal equilibrium. The linear relation at low-temperature limit shows the effective cooling of the electron temperature as low as 25 mK.

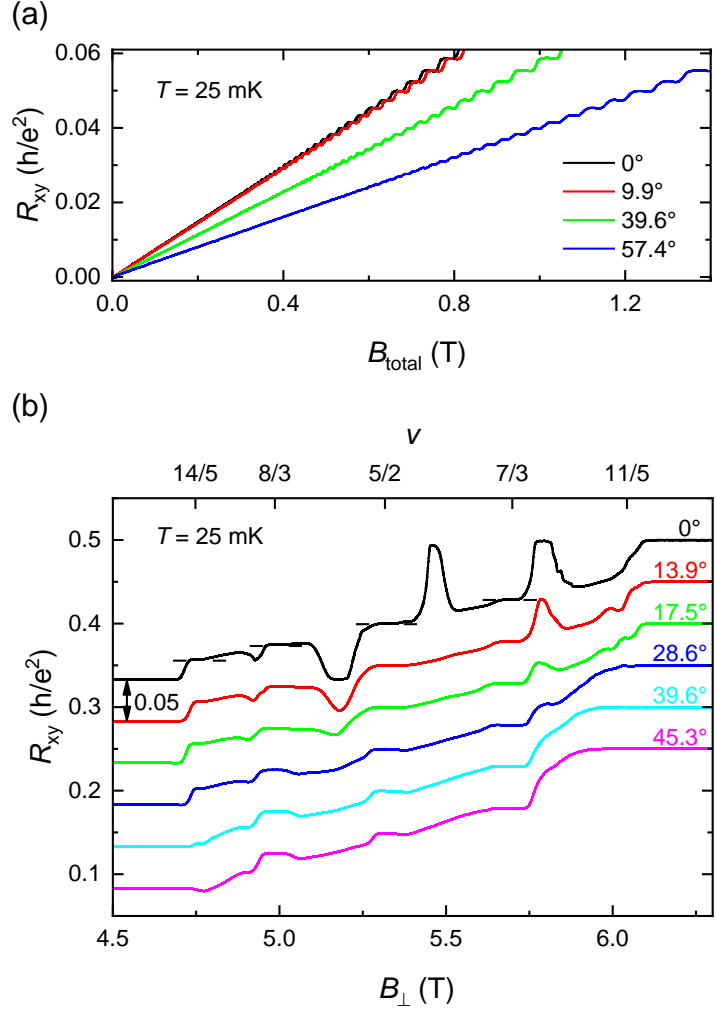

**Fig. 6** (a) Angle dependence of the Hall resistance as a function of total magnetic field $B_{total}$ at a probe temperature of 25 mK. The tilted angle is the angle between the applied magnetic field direction and the normal direction of the 2DEG, as in Fig. 2(c). (b) Angle dependence of the Hall resistance as a function of perpendicular magnetic field $B_\perp$ in the lower spin branch of the second Landau level ($2 < \nu < 3$) at a probe temperature of 25 mK. The traces with labeled tilted angles are shifted by an integer of 0.05 $h/e^2$ for clarity.